\documentclass[twocolumn,english,prl,showpacs]{revtex4}
\usepackage[colorlinks=true,urlcolor=blue,citecolor=blue,linkcolor=blue]{hyperref}
\usepackage[T1]{fontenc}
\usepackage[latin9]{inputenc}
\usepackage{amssymb}
\usepackage{graphicx}
\usepackage{amsmath,color}
\usepackage{mathrsfs}
\usepackage{float}
\usepackage{indentfirst}
\usepackage{babel}
\usepackage[sort&compress]{natbib}
\usepackage{color}

\begin{document}

\title{Topological Phase Transition in the Hofstadter-Hubbard Model}

\author{Lei Wang$^{1}$, Hsiang-Hsuan Hung$^{2}$ and Matthias Troyer$^{1}$}
\affiliation{$^{1}$Theoretische Physik, ETH Zurich, 8093 Zurich, Switzerland}
\affiliation{$^{2}$Department of Physics, University of Texas at Austin, Austin, Texas 78712, USA}

\begin{abstract}
We study the interplay between topological and conventional long range
order of attractive fermions in a time reversal symmetric
Hofstadter lattice using quantum Monte Carlo
simulations,  focussing on the case of one-third flux quantum per
plaquette. At half-filling, the system is unstable towards s-wave
pairing and charge-density-wave order at infinitesimally small
interactions. At one-third-filling, the noninteracting system is
a topological insulator, and a nonzero critical interaction strength is needed to
drive a transition from the quantum spin Hall insulator to a
superfluid. We  probe the topological signature of the
phase transition by threading a magnetic flux through a cylinder 
and observe quantized topological charge pumping. 
\end{abstract}


\maketitle
The Hofstadter model~\cite{Hofstadter:1976wt}, which describes electron moving
in a 2D lattice subject to a uniform magnetic field,  shows
an intriguing interplay between band structure and magnetic field
giving rise to a fractal energy spectrum and integer quantum Hall
states~\cite{Anonymous:TrbXgpDs,Osadchy:2001jm}. This model
has recently been realized in graphene superlattices~\cite{Dean:2013bv,
Hunt:2013ef, Ponomarenko:2013hlb} and  ultracold atoms in
optical lattices~\cite{Aidelsburger:2013ew, Miyake:2013jw}. The
 latter cold atom experiments, in particular, have realized  the Hofstadter model for {\em two}
spin components using {\em opposite} magnetic 
fluxes~\cite{Aidelsburger:2013ew,PhysRevLett.111.225301}. This system
conserves time-reversal-symmetry (TRS) and, when loading fermions into the optical lattice, is a natural realization of the quantum spin Hall (QSH) effect~\cite{Kane:2005hlb,Bernevig:2006ij} thus connecting the Hofstadter to the active field of topological insulators
(TI)~\cite{Hasan:2010ku,Qi:2011hb}. Being a topological insulating state in 2D, 
the QSH state is one of the first topologically insulating states observed in nature~\cite{Kane:2005hlb, Bernevig:2006ij,Konig:2007hs}. Tunable local interactions in cold atom experiments allow to address the  interesting interplay of interaction effects and the band topology. The problem has been studied in various models such as the Kane-Mele Hubbard model~\cite{Rachel:2010gqa,Hohenadler:2011kk,PhysRevB.85.115132,
PhysRevLett.107.010401,PhysRevB.84.205121,Assaad:2013fh},
the interacting Haldane
model~\cite{Varney:2010eja,Varney:2011jx,Anonymous:5IwlZp2x} and
the interacting Bernevig-Hughes-Zhang model~\cite{Wang:2012eq,
Yoshida:2013hua}, see~\cite{Hohenadler:2013dx} for a recent
review. 

There are, in general, two difficulties when studying interacting effect in topological insulators: the lack of unbiased numerical methods and the difficulty of direct quantification of the topological property of an interacting system. In this Letter we report on a large-scale quantum Monte Carlo
(QMC) study of the attractive Hofstadter-Hubbard model, where we overcome both these problems.  The sign problem ~\cite{Troyer:2005hv, Wu:2005im} is absent 
due to the time reversal symmetry of our model. We can thus map out the ground state phase
diagram and study the quantum phase transition from a QSH state to a superfluid upon increasing a local attraction between the two spin  species.
To directly diagnose the topological
nature of the correlated TI, we apply Laughlin's flux insertion technique~\cite{Laughlin:1981tm} and observe an induced topological charge pumping effect \cite{Anonymous:sQ6rtndT,
Anonymous:Apdv_csN}. 
These results provide direct evidence for a QSH state in an interacting system and a quantum phase transition to a topologically trivial superconductor.


\begin{figure}[tbp]
\centering
\includegraphics[width=9cm]{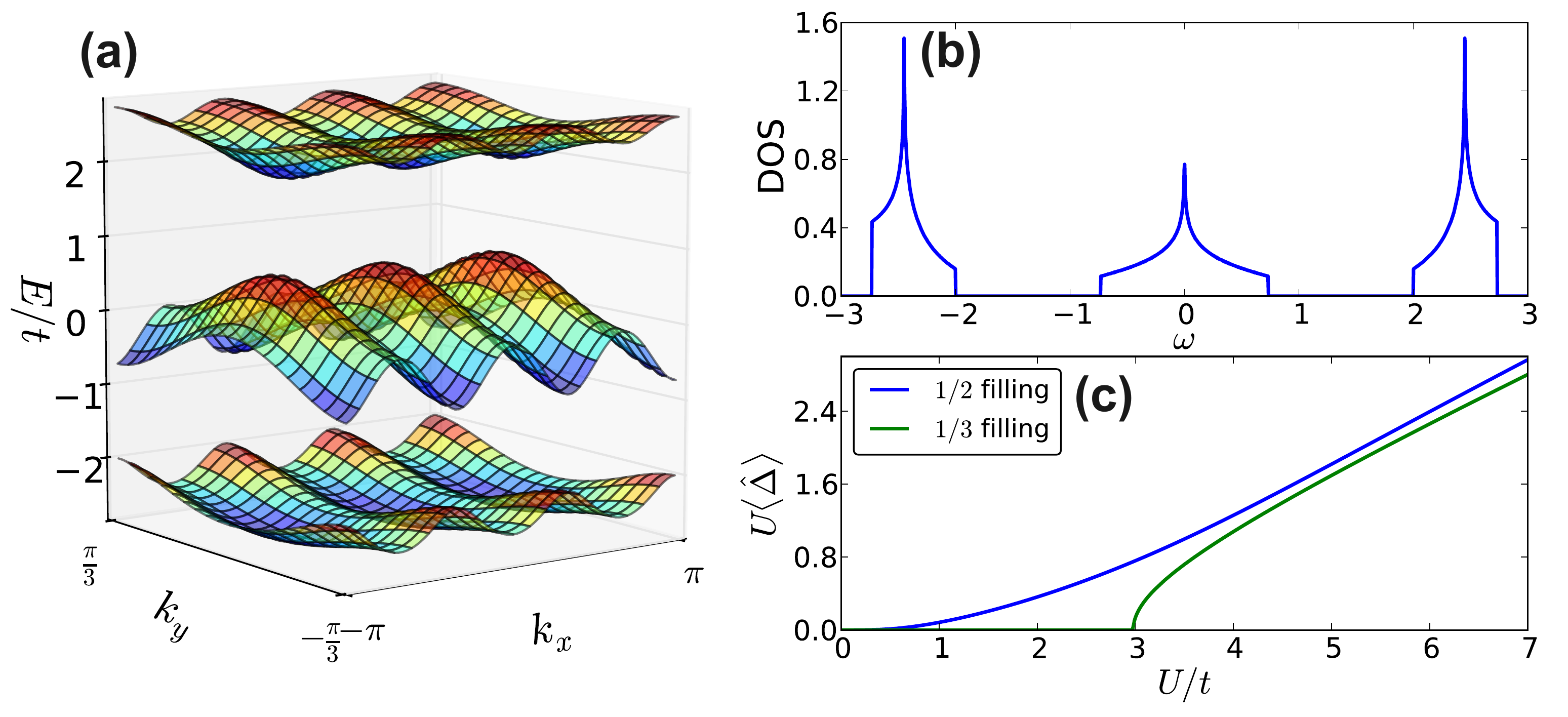}
\caption{(a) Band structure and (b) density of states of the noninteracting Hofstadter model with $\phi=1/3$. (c) Mean field result of the superconducting order parameter as a function of attractive interaction strength at $1/2$ and $1/3$ fillings.}
\label{fig:meanfield}
\end{figure}

\paragraph{The Model --}
The Hamiltonian of the time reversal symmetric Hofstadter-Hubbard model reads~\cite{Cocks:2012jj},
\begin{eqnarray}
\hat{H} & = & \hat{K} + \hat{V}, \nonumber \\
\hat{K} &= &-t\sum_{\mathbf{r},\sigma} e ^{i\sigma 2\pi y_{\mathbf{r}}\phi} \hat{c}^{\dagger}_{\mathbf{r}+\hat{\mathbf{x}} \sigma} \hat{c}_{\mathbf{r}\sigma}+\hat{c}^{\dagger}_{\mathbf{r}+\hat{\mathbf{y}}\sigma} \hat{c}_{\mathbf{r}\sigma} + h.c, \nonumber
\\ \hat{V}  &= & - U \sum_{\mathbf{r}} (\hat{n}_{\mathbf{r}\uparrow}-\frac{1}{2})(\hat{n}_{\mathbf{r}\downarrow}-\frac{1}{2}),
\label{eq:Ham}
\end{eqnarray}
where the operator $\hat{c}_{\mathbf{r}\sigma}$ annihilates a fermion of spin $\sigma=\pm1$ (corresponding to spin $\uparrow$ and $\downarrow$ fermions). at site $\mathbf{r}=(x_{\mathbf{r}},y_{\mathbf{r}})$. The phase factor $e^{i\sigma 2\pi y_{\mathbf{r}}\phi}$ in the
hopping amplitude  introduces magnetic flux $\pm\phi$ per plaquette for both spins. 
We will focus on the attractive case $- U< 0 $ with flux $\phi=1/3$, and on half and one third filling, while models with repulsive interactions and even denominators of $\phi$ have been studied in Refs.~\cite{Chang:2012fg,Cocks:2012jj}. In our case, the magnetic flux enlarges the unit cell threefold and there are
three energy bands, as shown in Fig.~\ref{fig:meanfield}(a-b). Since the hopping terms only connect different sublattices,
the spectrum preserves particle hole symmetry. At half-filling the noninteracting system is a metal with a nested Fermi surface.
A mean field treatment of the interactions shows an instability towards $s$-wave pairing  for infinitesimally
small attraction (see Fig.\ref{fig:meanfield}(c)). On the other
hand, the noninteracting system at $1/3$ filling is a topological
insulator, and mean field theory predicts a finite critical
interaction strength $U/t=2.95$ for a transition from a correlated quantum spin Hall insulator to a BCS state, because of a
vanishing density of states at Fermi level (see Fig.~\ref{fig:meanfield}(b)). 

\begin{figure}[!t]
\centering
\includegraphics[width=9cm]{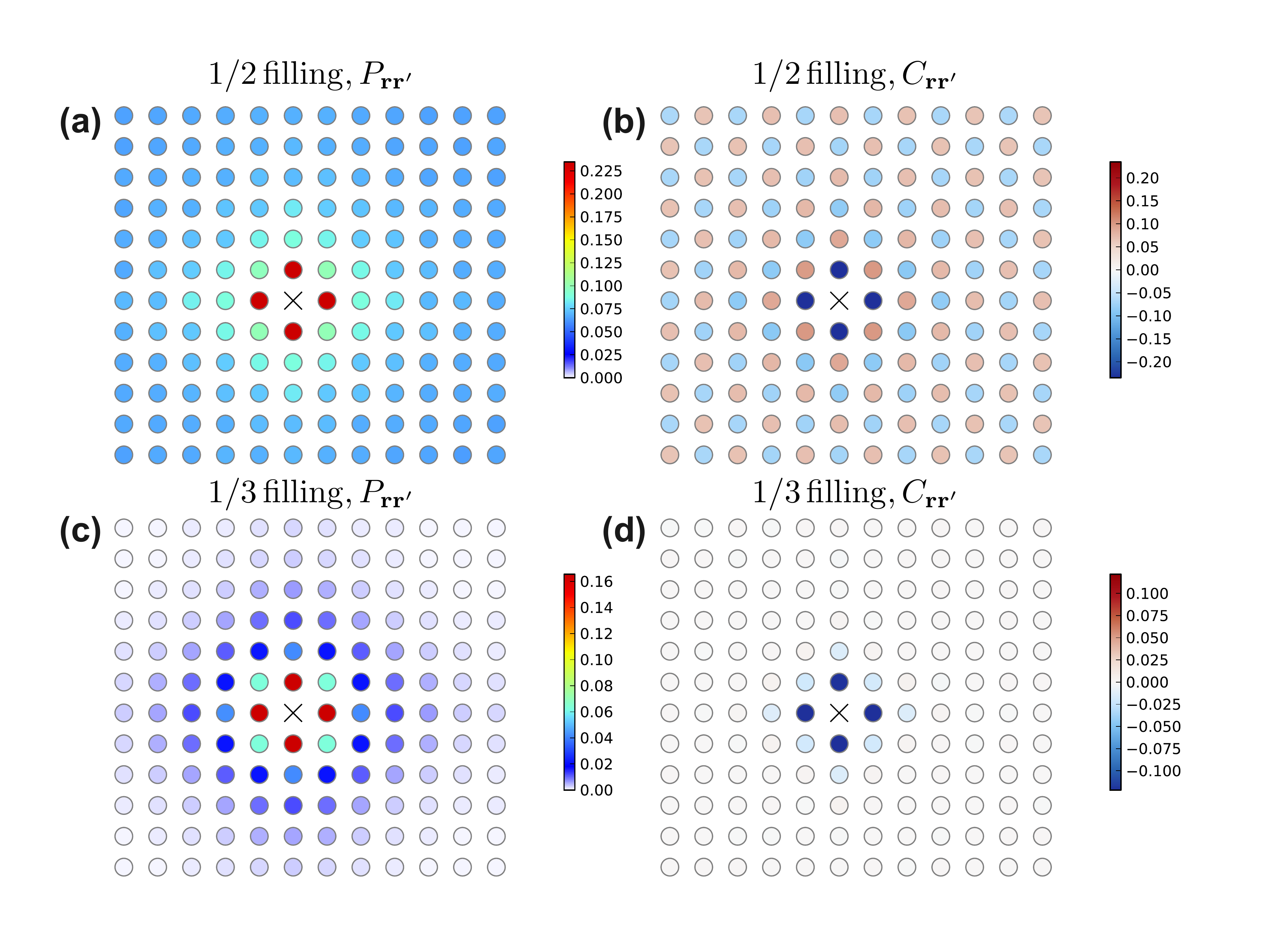}
\caption{Superconducting and density correlation functions at $U/t=4$. The $\times$ symbol denotes the site $\mathbf{r}$ relative to which correlations are shown.  (a) and (b) show results for the half-filled system, while (c) and (d) for the one-third-filled system. }
\label{fig:corr}
\end{figure}


\paragraph{Simulation method --}
We have simulated isotropic square lattices with linear size $L$ up to $24$ using an auxiliary field projective
QMC algorithm~\cite{Assaad:2008hx}, in which we calculate the ground state expectation values of
 observables $\hat{O}$ as 
\begin{equation}
\langle \hat{O} \rangle= \frac{\langle\Psi_{T}|e^{-\Theta \hat{H}/2} \hat{O} e^{-\Theta \hat{H}/2}|\Psi_{T}\rangle}{\langle\Psi_{T}|e^{-\Theta \hat{H}}|\Psi_{T}\rangle}.
\label{eq:projection}
\end{equation}
Using a trial wave function $|\Psi_{T}\rangle$ with non-vanishing overlap with the ground state, Eq.(\ref{eq:projection})  approaches  the ground state expectation value in the large $\Theta$ limit. We use the ground state of the noninteracting Hamiltonian $\hat{K}$ as trial state $|\Psi_{T}\rangle$ and $\Theta t=50$~\footnote{At half-filling to avoid open shell effect we add one additional flux quantum to the system~\cite{Assaad:2002de} when generating $|\Psi_{T}\rangle$.}. We break the projection into small steps and use the second-order Trotter-Suzuki decomposition
$e^{-\Delta \tau\hat{H}} = e^{-\Delta \tau\hat{K}/2} e^{-\Delta \tau\hat{V}} e^{-\Delta \tau\hat{K}/2} + O(\Delta \tau^{3})$
for each time step, where  $\Delta \tau t= 0.05$. The interacting term $e^{-\Delta \tau\hat{V}}$ is decomposed using the Hubbard-Stratonovich transformations
\begin{eqnarray}
  e^{ \Delta \tau U (\hat{n}_{\mathbf{r}\uparrow}-\frac{1}{2})(\hat{n}_{\mathbf{r}\downarrow}-\frac{1}{2}) } & =& \frac{ e^{- \Delta \tau
  U/4} }{2}\sum_{s = \pm 1} e^{\alpha s ( \hat{n}_{\mathbf{r}\uparrow} +
\hat{n}_{\mathbf{r}\downarrow} -1 )}  \label{eq:chargedecomp} \\
  & =&  \frac{e^{\Delta \tau U/4}}{2}
  \sum_{s= \pm 1} e^{i \gamma s ( \hat{n}_{\mathbf{r}\uparrow} -\hat{n}_{\mathbf{r}\downarrow} )} , \label{eq:magdecomp}
\end{eqnarray}
with $\cosh(\alpha) =  e^{\Delta \tau U/2}$ and $\cos(\gamma) =  e^{- \Delta \tau U/2}$. Both decompositions Eq.(\ref{eq:chargedecomp}-\ref{eq:magdecomp})  respect the time-reversal-symmetry and do not introduce sign problem in the Monte Carlo simulation. We use the decomposition Eq.(\ref{eq:chargedecomp}) for the one-third filled and Eq.(\ref{eq:magdecomp}) for the half-filled system.

\paragraph{Correlations --}
We first show the superconducting pair and the density correlation functions
\begin{eqnarray}
P_{\mathbf{r}\mathbf{r}^{\prime}} &=& \langle \hat{\Delta}^{\dagger}_{\mathbf{r}}  \hat{\Delta}_{\mathbf{r}^{\prime}} + \hat{\Delta}_{\mathbf{r}}  \hat{\Delta}^{\dagger}_{\mathbf{r}^{\prime}} \rangle,
\nonumber \\
C_{\mathbf{r}\mathbf{r}^{\prime}} &=& \langle\hat{n}_{\mathbf{r}} \hat{n}_{\mathbf{r}^{\prime}} \rangle - \langle\hat{n}_{\mathbf{r}} \rangle \langle \hat{n}_{\mathbf{r}^{\prime}} \rangle,
\label{eq:corr}
\end{eqnarray}
in Fig. \ref{fig:corr},
where $\hat{\Delta}^{\dagger}_{\mathbf{r}} =\hat{c}^{\dagger}_{\mathbf{r}\uparrow}\hat{c}^{\dagger}_{\mathbf{r}\downarrow}$ and $\hat{n}_{\mathbf{r}} = \hat{n}_{\mathbf{r}\uparrow} + \hat{n}_{\mathbf{r}\downarrow}$. At $U/t=4$ both superconducting and density correlations at half-filling extend to the farthest lattice site (in right corner), as shown in Fig.~\ref{fig:corr}(a) and (b) . Although the four-fold rotational symmetry of the square lattice is broken by the choice of Landau gauge in Eq.(\ref{eq:Ham}), it is restored in the  correlation functions. The density correlation shows a staggered pattern, indicating the tendency towards forming checkerboard charge-density-wave (CDW) order. Figure~\ref{fig:corr}(c) and (d) shows that at $1/3$ filling both the superconducting and CDW correlations are suppressed. In particular, the density correlation decays very rapidly. This suggests that  mean-field theory overestimates the extent of the ordered phase.

\begin{figure}[tbp]
\centering
\includegraphics[width=9cm]{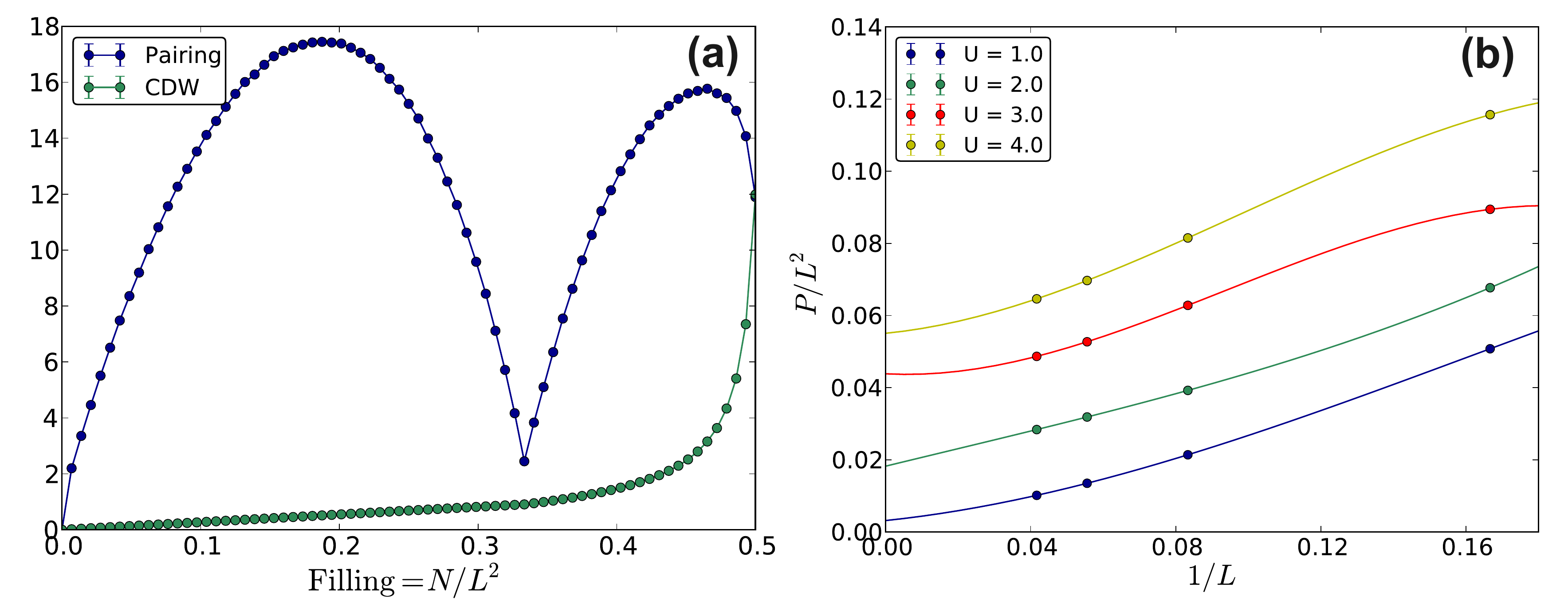}
\caption{(a) Superconducting and CDW structure factors  versus band filling in a $L=12$ lattice with $U/t=4$. (b) The superconducting structure factor versus $1/L$ for various interaction strengths at half-filling. The CDW structure factors are identical at half-filling due to a symmetry that is discussed in the text.}
\label{fig:halffilling}
\end{figure}

To better describe the interplay between the superconducting and CDW instabilities, we calculate their respective structure factors,
\begin{eqnarray}
P &=& \frac{1}{L^{2}} \sum_{\mathbf{r},\mathbf{r}^{\prime}} P_{\mathbf{r}\mathbf{r}^{\prime}}, \label{eq:P} \\
C &=& \frac{1}{L^{2}}\sum_{\mathbf{r},\mathbf{r}^{\prime}} e^{-i\mathbf{Q}(\mathbf{r}-\mathbf{r}^{\prime})}C_{\mathbf{r}\mathbf{r}^{\prime}},
\label{eq:C}
\end{eqnarray}
where $\mathbf{Q}=(\pi,\pi)$. Figure.~\ref{fig:halffilling}(a) shows the superconducting and
CDW structure factor versus fillings at $U/t=4$. The results
are symmetric around half-filling because of the particle-hole
symmetry of the model. The CDW structure factor
drops rapidly away from half-filling, indicating suppress of CDW
order away from commensuration filling. The superconducting structure factor
shows nonmonotonic behavior with filling factor. The variation is a
reminiscent of the shape of the density of states
(Fig.\ref{fig:meanfield}) and is large when the phase is metallic.
There is a pronounced minimum at $1/3$ filling when the lowest band
is fully filled, showing that a band insulating state strongly
disfavors forming of off-diagonal long range order.

At half-filling, the superconducting structure factor shows a
dip and becomes equal to the CDW structure factor due to an SU(2) symmetry. 
To reveal this symmetry 
we perform a particle hole transformation for the spin-down component
$\hat{c}_{\mathbf{r}\downarrow}\rightarrow e^{-i\mathbf{Q}\mathbf{r}}
\hat{c}^{\dagger}_{\mathbf{r}\downarrow}$. This transformation
reverses both the sign of interaction $U$ and the magnetic flux
$\phi$ of spin down particles. The resulting repulsive model has
same magnetic flux for both spin components and is manifestly
SU(2) symmetric~\footnote{This is different from the
Kane-Mele-Hubbard model~\cite{Hohenadler:2011kk} and the model
studied in~\cite{Cocks:2012jj}, where SU(2) spin symmetry is broken
explicitly.}, explaining why superconducting and pair correlations are identical. The model is in a supersolid  phase like the attractive Hubbard model at half filling~\cite{Hirsch:1985wz}. 
The
 reason for this degeneracy is also transparent in the strong
coupling limit, where the attractive interaction binds two
fermions with opposite spin into (spinless) hard core bosons, which are inert to
magnetic flux. The resulting model lies exactly at the
Heisenberg point~\cite{Schmid:2002fu} and shows degenerate 
CDW and superfluidity instabilities. 

\paragraph{Determining the transition --}
Figure~\ref{fig:halffilling}(b) shows
an extrapolation of $P/L^{2}$ at half filling (equals to $C/L^{2}$)
using a third order polynomial in $1/L$. The extrapolated value gives
the square of the pairing order parameter~\cite{Huse:1988wh,
Scalettar:1989fi}. Our results show that at half filling the system is already ordered at weak
coupling $U/t=1$, in agreement with the mean-field results.

\begin{figure}[tbp]
\centering
\includegraphics[width=9cm]{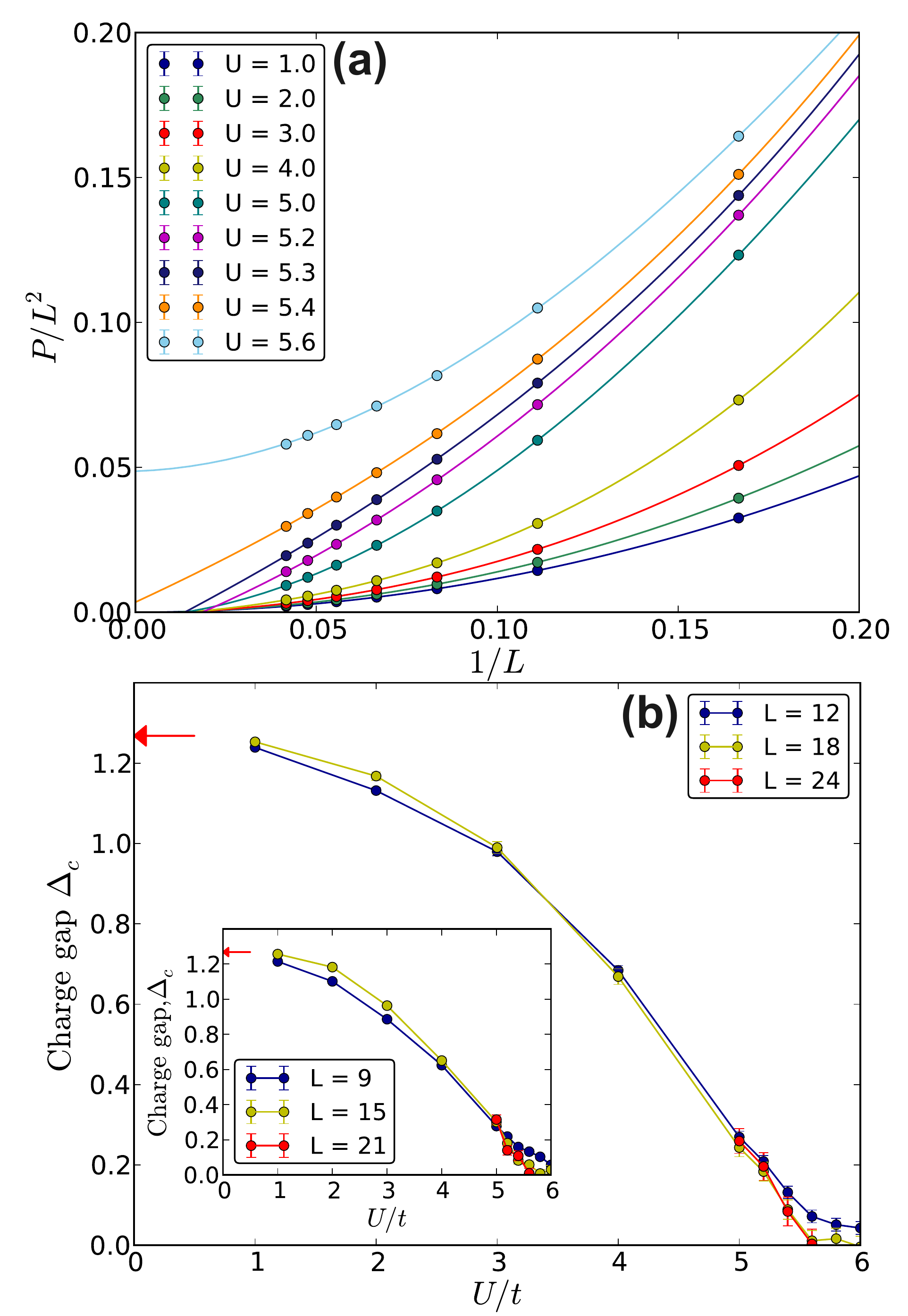}
\caption{(a) The superconducting structure factor of a $1/3$-filled system. Solid lines are fits with cubic function of $1/L$. (b) The charge gap $\Delta_c$ versus  interaction strength $U$ for systems with periodic boundary conditions. The inset shows the charge gap in systems with anti-periodic boundary conditions. The red arrow indicates the band gap of the noninteracting system.}
\label{fig:qshe}
\end{figure}

However, as we have shown before, mean-field theory erroneously
predicts a supercondunctiong state at $U/t=4$ and one third
filling~\footnote{The CDW correlation is substantially suppressed at
this filling and does not compete with superconductivity.}. 
Figure.~\ref{fig:qshe}(a) shows the superconducting structure
factor for various interaction strengths at $1/3$ filling.
Extrapolation indicates that the transition to a superconducting state happens at
$U/t\gtrsim 5.4$,  a substantially larger value than the mean-field prediction $U/t=2.95$ (Fig.\ref{fig:meanfield}(c)). Since the time-reversal symmetry is conserved, there is no vortex lattice structure in the BCS state compare to the model studied in Ref.~\cite{Zhai:2010iaa}. Again, this agrees with the fact that the hard core bosons in the strong interaction limit are inert to magnetic field.

An alternative determination of the critical interaction strength for the QSH to BCS is based on the charge gap, which can be calculated as
\begin{equation}
\Delta_{c} = \frac{E_{L^{2}/3+1} + E_{L^{2}/3-1} -2E_{L^{2}/3}} {2},
\label{eq:chargegap}
\end{equation}
where $E_{N}$ is the ground state energy of $N_{\uparrow}
=N_{\downarrow}= N$ particles. $\Delta_{c}$ is the 
energy cost of adding a pair of fermions to the system. In the noninteracting limit it equals to
the minimal band gap $1.268t$, realized at the momenta $(\pm\frac{\pi}{3},
\frac{\pi}{3})$ and $ (\pi, \frac{\pi}{3})$ in the Brillouin zone (see. Fig.~\ref{fig:meanfield}(a)). We choose boundary conditions of finite size clusters carefully to ensure that these momenta exist: we use periodic boundary condition for 
$L=6,12,18,24$ and anti-periodic boundary condition for $L=3,9,15,21$. Figure~\ref{fig:qshe}(b) shows the charge gap
versus $U$ for various system sizes. It decreases from the
noninteracting value as the attractive interaction
increases, and becomes zero at the quantum phase transition to the superconducting phase. We find that this happens at $U/t=5.6$ for
the largest system we have calculated, consistent with transition
point $U\gtrsim5.4$ estimated from the superconducting correlations.

%

\begin{figure}[tbp]
\centering
  \includegraphics[width=9cm]{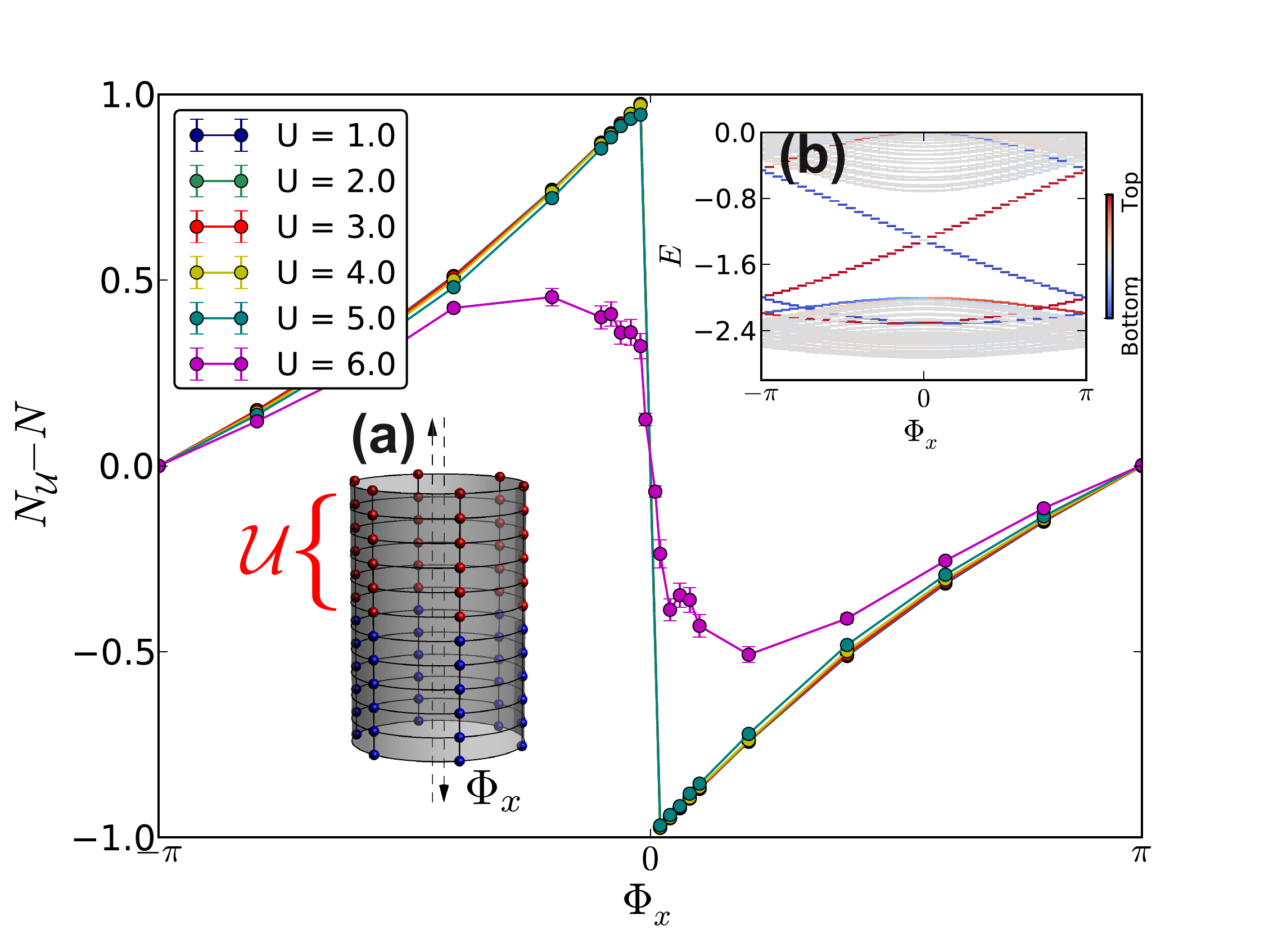}
\caption{Total particle number in the upper-half of a cylinder
(Eq.(\ref{eq:NU})) versus flux $\Phi_{x}$ for various
interaction strengths. Inset (a) illustrates a flux $\Phi_{x}$ (opposite for spin
$\uparrow$ and $\downarrow$) threaded through the cylinder, which pumps particles
vertically in the QSH phase. Sites in the upper half of the cylinder are shown in red and their occupation numbers sum to $N_{\mathcal{U}}$. 
Inset (b) shows single particle energy
levels versus the threaded flux $\Phi_{x}$ in a cylindrical 
noninteracting system. Color indicates the center-of-mass of
single particle wave functions, with red (blue) being closer to the
upper (lower) edge. } \label{fig:pumping}
\end{figure}

\paragraph{Topological nature of the transition --}

The QSH state is
characterized by a nontrivial topological $\mathbb{Z}_{2}$
index~\cite{Kane:2005gb}, and we expect the systen to remain in this phase until the
transition point. Several approaches have been proposed to
characterize interacting topological
insulators~\cite{Wang:2011jh, Wang:2012hj, Wang:2012fj}. However
they either requires approximations~\cite{Wang:2012eq} or miss the
interaction induced phase transition~\cite{Hung:2013hg, Lang:2013em,
Hung:2013tb, Meng:2013id}.

To directly reveal the topological nature of the QSH-BCS transition, we 
thread a flux through a cylinder~\cite{Laughlin:1981tm} and probe
the topological charge pumping
effect~\cite{Anonymous:sQ6rtndT,Anonymous:Apdv_csN,
Alexandradinata:2011bo}. Figure~\ref{fig:pumping}(a) shows a
flux $+\Phi_{x}$ ($-\Phi_{x}$) for spin up  (down)
particles threaded through a cylinder, which amounts to introducing spin-dependent twisted
boundary conditions~\cite{Qi:2006co, Sheng:2006bra}. Since spin up
and down particles feel opposite magnetic fluxes (both $\Phi_{x}$
and $\phi$), time reversal symmetry is preserved and there is no sign problem in the
QMC simulations. When $\Phi_{x}$ changes from
$-\pi$ to $\pi$ both spin up and  down particles are pumped along
the same vertical direction~\footnote{This is different from the
spin pump studied in Ref.~\cite{Fu:2006jk} where up and down
particles move in opposite directions.}. The total pumped charge is
proportional to the spin Chern number~\cite{Sheng:2006bra}, which is the difference of the Chern number for spin up and
 down particles and directly probes the quantum spin Hall effect.

To get a better understanding of the topological pumping effect, 
 inset (b) of Fig.~\ref{fig:pumping} shows the single
particle energy spectrum versus $\Phi_{x}$ on a cylinder with a circumference of six sites. Color indicates the
center-of-mass position of each eigenstate, with red (blue) color
being closer to the upper (lower) edge. There are two edge states
corresponding to the cylinder's top and bottom edge, which cross at
$\Phi_{x}=0$.  In the $1/3$ filled system, the Fermi level lies
exactly at this crossing point and the density distribution is
symmetric in the absence of a  flux $\Phi_{x}$. Inserting an
infinitesimal flux moves the particles towards to one of the edges and introduces
a polarization along the cylinder. Further increase of
$\Phi_{x}$ pumps the particle vertically, thereby changing the polarization.
The total change of polarization upon inserting a $2\pi$  flux gives
the spin Chern number. This topological pumping effect is robust
against interactions and can be used to distinguish the
correlated $\mathbb{Z}_{2}$ topological insulators and a
topological trivial superfluid state.

To quantify the topological pumping effect we calculate the total particle number in the upper part of the cylinder (see Fig.\ref{fig:pumping}(a))
\begin{equation}
N_{\mathcal{U}} = \sum_{\mathbf{r}\in \mathcal{U}} \left( n_{\mathbf{r}\uparrow} + n_{\mathbf{r}\downarrow} \right ).
\label{eq:NU}
\end{equation}
The total number of particles pumped to the upper half of the cylinder is
$\int_{-\pi}^{\pi}\frac{dN_{\mathcal{U}}}{d\Phi_{x}} d\Phi_{x}$.
Because $N_{\mathcal{U}}$ is periodic with $\Phi_{x}$, there must be
discontinuities in $N_{\mathcal{U}}$ to account for the finite
shift. The size of the discontinuity is again equal to the spin-Chern
number, and strong  evidence for the presence of a QSH state in the interacting system. Figure \ref{fig:pumping} shows $N_{\mathcal{U}}-N$
versus $\Phi_{x}$ for various interaction
strengths~\footnote{Because of the presence of low lying excited
states close to $\Phi_{x}=0$, we use much larger projection
parameter $\Theta t=200$ to obtain a converged ground state density.}.
Inside the QSH state ($U/t=1,2,3,4,5$), the curves are almost identical and they all
show an overall shift of two, which is the spin Chern number
of the QSH state. The discontinuity at $\Phi_{x}=0$ is a characteristic
behavior of the topological nontrivial state.
After the transition to the BCS state ($U/t=6$) the discontinuity
disappears and the pumped charge is zero.
These results provide a direct
topological signature of the QSH-BCS transition. Being related but
different from the $\pi$-flux insertion method~\cite{Ran:2008dva,
Qi:2008kza} used in Ref.~\cite{Assaad:2013fh}, the charge pumping
approach directly probes the topological response of a QSH state
and can be easily generalized to systems with spin-flip
terms~\cite{Sheng:2006bra, Qi:2008kza, Alexandradinata:2011bo}.





\paragraph{Outlook --}
Our work opens up a number of exciting possibilities
for studying strongly correlated topological phases using numerical
exact quantum Monte Carlo methods. For example, it will be interesting to
further study the correlation effect in a Hofstadter model with
arbitrary fluxes, where the occupied band has higher Chern
number or even a fractal energy spectrum.  Topological charge pumping probe can also be used to 
identify the fractional topological phases~\cite{PhysRevLett.100.030404, KunYang:2012}. 
A detailed study of edge physics in conjunction with the topological pumping probe will also be of
 interest. Experimentally, our results are directly elevant to current studies
of Hofstadter model in cold atom
systems~\cite{Chin:2013kc,Aidelsburger:2013ew,Miyake:2013jw}. Along
this line, we leave a detailed study of the finite temperature
phase diagram and inhomogeneity effects for future study.

\paragraph{Acknowledgments --}
We thank J. Gukelberger, M. Dolfi  and A. Soluyanov for discussion and support. 
Simulations were performed on the M\"{o}nch cluster of
Platform for Advanced Scientific Computing (PASC), the Brutus
cluster at ETH Zurich, and the ``Monte Rosa'' Cray XE6 at the Swiss
National Supercomputing Centre (CSCS).  H.H.H. acknowledges the
computational support from the Center for Scientific Computing at
the CNSI and MRL through NSF MRSEC (DMR-1121053) and NSF CNS-0960316.
We have used ALPS libraries~\cite{BBauer:2011tz} for Monte Carlo
simulations and data analysis. This work was supported by ERC grant SIMCOFE.

\bibliographystyle{apsrev4-1}
\bibliography{HofstadterQMC}

\end{document}